\documentclass[useAMS,usenatbib]{mn2e}

\input{epsf}
\usepackage{natbib}
\usepackage{graphicx}	
\usepackage{amsmath}

\usepackage{journalabbreviations}
\newcommand{\nodata}{ ~$\cdots$~ }%

\def\Psgs{Post-starburst galaxies}
\def\psgs{post-starburst galaxies}
\def\psg{post-starburst galaxy}

\makeatletter

\newcommand{\Rmnum}[1]{\expandafter\@slowromancap\romannumeral #1@}
\makeatother
\newcommand\ion[2]{#1$\;${\small\Rmnum{#2}}\relax}%

\title{Post-Starburst Quasars: Bridging the Gap Between Post-Starburst Galaxies and Quasars}
\author[S. L.Cales and M. S. Brotherton]{Sabrina L. Cales$^{1,2}$, and Michael S. Brotherton$^{3}$\\$^{1}$Yale Center for Astronomy and Astrophysics, Physics Department, Yale University, New Haven, CT, 06511; sabrina.cales@yale.edu\\ 
$^{2}$ Department of Astronomy, University of Concepcion, Concepcion, Chile\\
$^{3}$Department of Physics and Astronomy, University of Wyoming, Laramie, WY 82071; mbrother@uwyo.edu}

\begin{document}





\date{Received 2014 July 21; in original form 2013 July 21}

\pagerange{\pageref{firstpage}--\pageref{lastpage}} \pubyear{2014}

\maketitle

\label{firstpage}


\begin{abstract}

In order to better understand the nature of post-starburst quasars (PSQs) in the context of galaxy evolution, we compare their properties to those of post-starburst galaxies and quasars from appropriately selected samples possessing similar redshift ($z \sim 0.3$), luminosity ($M_r \sim -$23), and data quality.  We consider morphologies, spectral features, and derived physical properties of the stellar populations and central supermassive black hole.  PSQs themselves come in two types: the more luminous AGNs with more luminous post-starburst stellar populations hosted by elliptical galaxies, some which are clearly merger products, and the less luminous systems existing within relatively undisturbed spiral galaxies and possessing signs of a more extended period of star formation.  Post-starburst galaxies (PSQs) have elliptical and disturbed/post-merger morphologies similar to those of the more luminous PSQs, display similar spectral properties, but also can have younger stellar populations for a given starburst mass.  Quasars at similar redshifts and luminosities around the Seyfert/quasar transition possess similar AGN characteristics, in terms of black hole mass and accretion rate, compared with those of PSQs, but do not appear to be hosted by galaxies with significant post-starburst populations. Recent studies of more luminous quasars find hosts consistent with those of our luminous PSQs, suggesting that these PSQs may be in transition between post-starburst galaxies and a more luminous quasar stage when the post-starburst stellar population remains dominant.  The lower luminosity PSQs appear to differ from lower luminosity quasars (Seyfert galaxies) in terms of more significant star formation in their past.

\end{abstract}

\begin{keywords}
galaxies: quasars, active, evolution
\end{keywords}

\section{Introduction}
\label{sec:Intro}

In major-merger driven evolutionary scenarios, dynamical interactions between merging galaxies drive gas in towards nuclear regions, which can activate a burst of star-formation and AGN activity. In due course, the AGN can also provide radiative or mechanical feedback truncating star-formation and extinguishing the AGN (\citealt*{dimatteo05}; \citealt*{springel05}; \citealt{hopkins06}). Major-merger driven evolution predicts the existence of galaxies undergoing relatively short-lived transitions ($\la$1Gyr), including (ultra-)luminous infrared galaxies, \psgs, post-starburst quasars (PSQs), and quasars. 

Various observations of different galaxy types lend support for this evolutionary scenario. Ultraluminous infrared galaxies (ULIRGs) have been found to be ubiquitously strong, interacting merger systems heated by star-forming and/or AGN power sources \citep[e.g.,][]{genzel98}. They have been hypothesized to evolve into normal quasars after the central engine clears away the dust associated with the massive burst of star formation \citep{sanders88, sandersmirabel96, veilleux06}. These objects are possibly the first observational snapshot in merger evolution and formation of a larger elliptical galaxy.

At least in some cases, the temporal evolution of a ULIRG may produce quasars that show contemporaneous post-starburst stellar and AGN signatures in their spectrum. Post-starburst quasars (PSQs) are broad-lined active galactic nuclei (AGNs) that possess the spectral signatures of massive ($M_{burst} \sim10^{10} M_{\odot}$), moderate-aged stellar populations (hundreds of Myr). Since the discovery of an extreme example, UN J1025$-$0040 \citep{brotherton99}, which is morphologically disturbed and has a companion galaxy also showing dominant post-starburst stellar signatures in its spectrum \citep{brotherton02, canalizo00}, large quasar surveys have permitted the identification of PSQs in much larger numbers.

\citet[][hereafter C11]{cales11} analyzed \textit{Hubble Space Telescope} (\textit{HST}) images of a sample of PSQs with redshifts between 0.25 and 0.4. They found a heterogeneous population with a variety of morphologies including host galaxies that are early-type, spiral, disturbed, interacting and isolated. Furthermore, spectral modeling of PSQs with early-type hosts have higher AGN luminosities ($\langle L_{Tot}\rangle \sim$ 10$^{44.08}$ erg s$^{-1}$) and younger, bursty stellar populations ($\langle$SB Age$\rangle \sim$960 Myr), while the spiral hosts have a more complex and extended star formation history \citep[][hereafter C13]{cales13}. The higher luminosity early-type PSQs with $L_{Tot} >$ 10$^{43.85}$ erg s$^{-1}$ appear to be the product of major-mergers, while the lower luminosity PSQs ($L_{Tot} <$ 10$^{43.95}$ erg s$^{-1}$) hosted in spirals are consistent with being secularly evolving Seyfert galaxies. Thus, at least in the case of these PSQs, both dynamical interactions and secular evolution appear to play important roles (i.e., elliptical mode and spiral mode in C13).

These studies support current ideas about mutual black hole/galaxy evolution in general, in that only the most luminous AGN (quasars, $M_R \leq -$22) seem to be triggered by, even require, major mergers \citep{hopkinshernquist09,treister12}. Furthermore, the lower luminosity AGN generally do not require mergers for triggering, have less massive black holes fueled at lower accretion rates and show no enhancement in disturbance fractions when compared to non-active galaxies \citep{cisternas11}. The number density of secularly evolving galaxies is thought to have  remained relatively constant over time and has only recently ($z\la1$) become the dominant mode of evolution \citep[e.g.,][]{heckmankauffmann11, alexanderhickox12}. Thus, the lower mass systems ($M_{BH} \la 10^{7.5} M_{\odot}$ and $M_{spheroid} \la 10^{10.5} M_{\odot}$) in the local universe preferentially show activity \citep[i.e., AGN cosmic downsizing, see][]{cowie96,heckman04}. The more luminous PSQs resulting from major mergers may represent low-redshift counterparts to the high-luminosity quasars much more common at redshifts of 2-3.

In C11 and C13 we utilized AGN/host galaxy light decomposition analysis of high quality imaging and spectroscopic data to determine PSQ morphology and AGN and post-starburst fundamental physical properties. Our goal in the current investigation is to increase our understanding of the C11 and C13 samples of PSQs, and to determine to what extent their properties are similar to, or differ from, post-starburst galaxies and AGN of similar redshift and luminosity.  We examine host morphologies, stellar population ages and masses, as well as black hole masses and their accretion rates.  We want to understand if PSQs are consistent with being transitionary objects between these other classes, or if they display fundamental similarities or differences that can inform our understanding of how these classes fit together.  

In \S 2, we describe the selection and properties of our PSQ sample, as well as several samples of post-starburst galaxies and quasars selected for comparison. In \S 3, we compare in quantitative detail for the first time appropriately matched samples, their morphological classifications and stellar populations. For consistency among samples we make our own morphological classifications of the post-starburst and low luminosity quasars. We fit post-starburst galaxy spectra with models of stellar populations consistent with the approach of C13 to obtain the ages and masses of the post-starburst and compare these to PSQs. In \S 4, we create composite spectra of our different classes of objects to illustrate their similarities and differences. In \S 5, we make quantitative and statistical comparisons of morphology, stellar populations and AGN properties among the different samples, and also the stellar populations of more luminous quasars, discuss the results and how PSQs may fit into the grand scheme of things. \S 6 presents our conclusions.

\begin{table*}\scriptsize
\begin{minipage}{140mm}
\caption{Sample Properties \label{tab:morphch4} }
\begin{tabular}{llccclc}
\hline \hline 
Sample & Reference\footnote{C11: \citep{cales11}, Y08: \citep{yang08}, G07: \citet{goto07}, W06: \citep{woo06}, T07: \citep{treu07}, Be10: \citep{bennert10}} & Number of & $z$ & $\langle M_r\rangle \pm \sigma_{mean}$ & Morphology & Fraction \\
 &  & Objects & Range &  &  & Disturbed \\
\hline
PSQs & C11 & 29 Imag., 38 Spec. & 0.25:0.45 & $-$22.89 $\pm$ 0.04 & 13 spiral, 13 early-type, 3 Indeterm. & 0.59 $\pm$ 0.09 \\
Post-starburst Galaxies & Y08 & 21 Imag. & 0.07:0.18 & $-$20.07 $\pm$ 0.02 & 7 spiral, 8 early-type, 6 Indeterm. & 0.52 $\pm$ 0.11\\
Post-starburst Galaxies & G07 & 28 Spec. & 0.25:0.45 & $-$23.22 $\pm$ 0.04 & \nodata & \nodata \\
$z \sim 0.36$ low luminosity quasars & W06, T07, Be10 & 20 Imag. \& Spec. & $z \sim 0.37$ & $-$23.25 $\pm$ 0.05 & 12 spiral, 6 early-type, 2 Indeterm. & 0.35 $\pm$ 0.11 \\
\hline
\end{tabular}
\end{minipage}
\end{table*}

\begin{figure}
\includegraphics[width=7.5cm,angle=270]{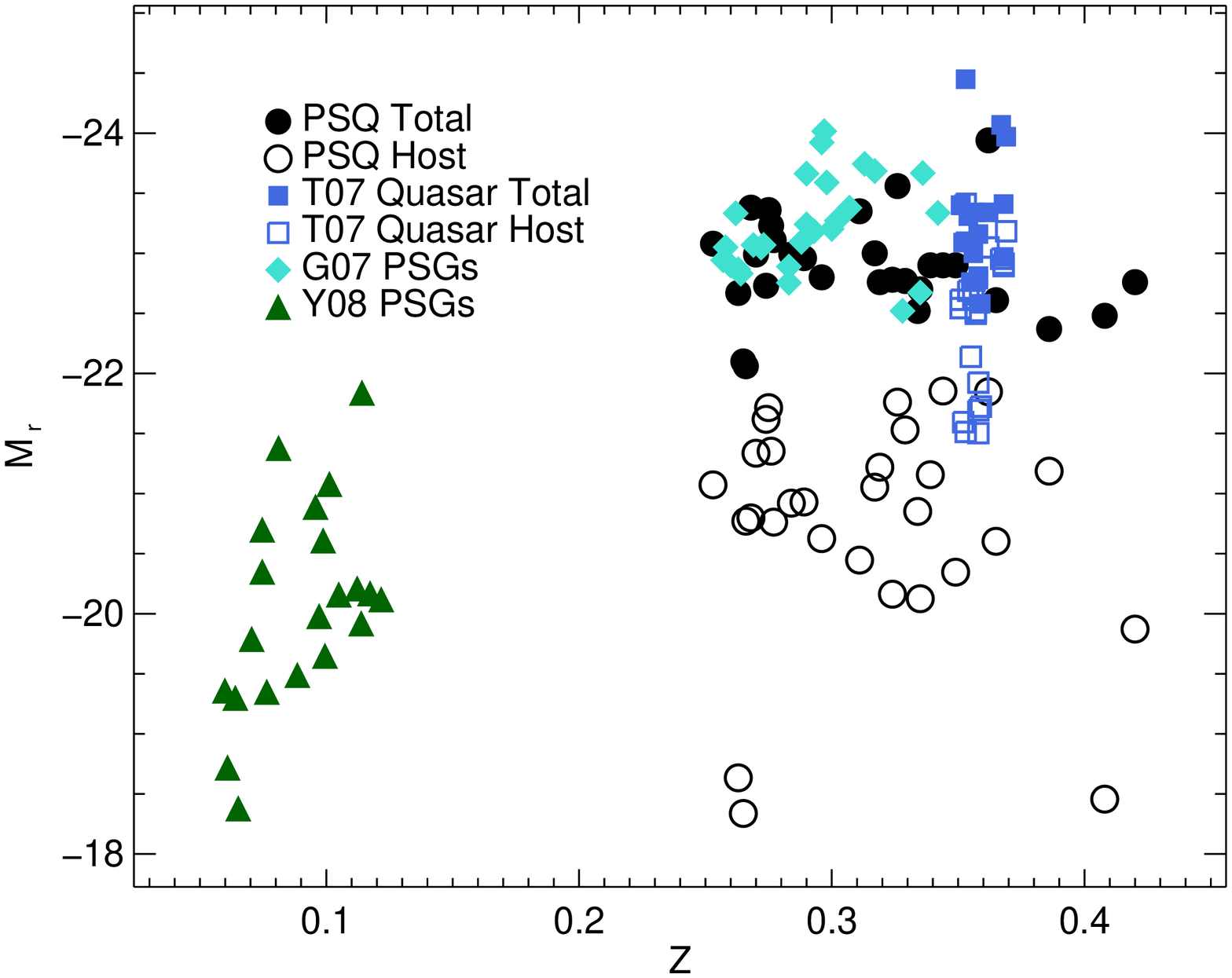}
\caption{Comparison of Galactic dereddened $K$-corrected $r$ absolute magnitudes of our PSQ sample (black circles), the quasars of \citet[][T07, blue squares]{treu07}, and the \psgs\ of \citet[][Y08, green triangles]{yang08} and  \citet[][G07, cyan diamonds]{goto07}. Filled symbols represent the total magnitude while the open symbols use $f_{nuc}$ from the original sources to calculate the host galaxy magnitude. The G07 spectroscopic sample and the T07 imaging and spectroscopic samples are well matched in redshift and absolute $r$ magnitude. However, due to a lack of high-quality imaging of the G07 sample we use HST-determined morphologies of the Y08 lower redshift sample as proxies for that of G07. Errors for redshift and absolute magnitude are smaller than the data points ($z_{err}$ $<$ 0.0002 and $Mr_{err}$ $<$ 0.03).}
\label{fig:compch4} 
\end{figure}

\section{Comparison Samples}
\label{sec:Samp}

We compare the morphologies and other derived properties of the C11 and C13 PSQs to post-starburst galaxies and broad-lined AGN of similar redshift and luminosity.  We use samples of objects with relatively unbiased selection, high-resolution imaging, and host galaxy spectral modeling consistent with the methods used for PSQs. There is an appropriate and well-studied sample of low luminosity quasars at a redshift of 0.36 \citep{treu07}, and although detailed spectral modeling is unavailable we can reach some conclusions based on composite spectra and will later consider results from higher luminosity quasars and their hosts. For post-starburst galaxies, we consider two samples. The first, \citet{yang08}, possesses HST imaging, which we utilized to make morphological classifications and to measure disturbance fractions. The objects are at lower redshift and luminosity than desirable, however the results are likely applicable to higher luminosity systems (see \S~\ref{sec:Res.Morph}). The second sample we draw from the catalog of \citet{goto07} and perform spectral modeling in order to measure fundamental parameters of the post-starburst (age and mass). Table~\ref{tab:morphch4} characterizes some observational properties of the samples while Figure \ref{fig:compch4} shows how well matched the samples are in $M_r-$redshift space. In the following subsections we provide more details about each sample and their available data. We begin by describing the sample selection of PSQs from C11 and then detail the selection of new comparison samples of \psgs\ and quasars.


\subsection{Post-Starburst Quasars}
\label{sec:Samp.PSQ}

Detailed descriptions of the PSQ sample selection can be found in C11. For convenience, we briefly describe the PSQ sample selection here. We selected PSQs having contemporaneous broad-lined AGN and post-starburst stellar populations by searching the SDSS data release 3 (DR3) online database and making measurements of the available spectra. Our selection criteria were:

\begin{itemize}
\item Broad emission lines as defined by the SDSS DR3 online database (FWHM $> 1000$ km s$^{-1}$)
\item $r$ model magnitudes $\la19$ 
\item 0.25 $< z < $0.45, in order to estimate black hole masses from the FWHM of Mg II, H$\beta$ and/or  H$\alpha$
\item $S/N > 8$ between rest-wavelengths of 4150 and 4250~\AA\ in the SDSS spectra
\item The sum of the Balmer absorption lines (H$\delta$, H$\zeta$, and H$\eta$) $>$~2~\AA\ rest-frame equivalent width at a significance greater than 6$\sigma$
\item H$\delta > 1$~\AA\  rest-frame equivalent width
\item Balmer Break $> 0.9$; defined by the ratio of the fluxes at two 100~\AA\ wide regions centered at rest-wavelengths 4035 and 3790~\AA
\end{itemize}

The catalog was visually inspected in order to reject spurious objects. We note that the selection was done prior to any decomposition so the quasar contribution to the total luminosity may not exceed a formal quasar luminosity cut and include formally post-starburst quasars as well as post-starburst Seyfert galaxies.  We refer to all objects selected above simply as PSQs.

Our images come from an HST F606W snapshot survey using the Advanced Camera for Surveys (ACS) targeting 29 PSQs (C11). SDSS spectra were enhanced with additional long-slit spectra using the Kitt Peak National Observatory (KPNO) 4 m Mayall telescope  or the Low-Resolution Imaging Spectrometer \cite[LRIS;][]{oke95} on the Keck I telescope. See C11 and C13 for further details regarding observations and data reductions.

\subsection{Post-Starburst Galaxies}
\label{sec:Samp.PSG}

Post-starburst galaxies have spectra that display a strong Balmer break and strong Balmer absorption lines indicative of A stars that are dominant in intermediate-age stellar populations, along with evidence that ongoing star formation is no longer significant \citep[e.g. weak \ion{O}{2} $\lambda$3727][]{dresslergunn83}. 

For a sample of \psgs\ matched in redshift and luminosity to the PSQs, we used the catalog of  \citet[][hereafter G07]{goto07} from the SDSS Date Release 5. G07 identifies extended objects - that are not classified spectroscopically as a star - with S/N greater than 10 (in r-band), strong post-starburst stellar absorption features indicated by $H_{\delta}$ EW~$>$~5.0~\AA, and little nebular emission ([\ion{O}{2}] EW~$>~-$2.5~\AA\ and $H_{\alpha}$ EW~$>~-$3.0~\AA) as \psgs. We matched this sample with our PSQ sample by comparing only those \psgs\ with $r$ model magnitudes $\la19$ and 0.25~$< z <$~0.45 (noting that the highest redshift post-starburst galaxy in the G07 catalog is 0.3421 and the faintest $r$ magnitude is $\sim18.6$). The resultant subsample is 28 \psgs.  We note that these selection criteria differ from those of the PSQs in that the Balmer absorption line equivalent widths are  always uniformly larger, but there is no quasar continuum to dilute these in \psgs, and the PSQs were allowed to have significant [\ion{O}{2}] emission.  Because of the latter, it may only be appropriate to  compare the G07 sample with our PSQs with weak [\ion{O}{2}] emission.


SDSS images of the \psgs\ at these redshifts are insufficient to determine morphologies or information about disturbance level.  There is a sample of \psgs\ with HST imaging, however, that is likely suitable.  High-resolution HST/ACS and WFPC2 images exist for a sample \psgs\ \citep[hereafter Y08]{yang08}. The selection criteria for this sample is:~average equivalent width H$_{\beta,\gamma,\delta}$~$>$~5.5~\AA\ and [\ion{O}{2}] EW~$>$~-2.5~\AA. We note that the redshift for the Y08 sample at 0.07~$< z <$~0.18 is  lower than the PSQs. However, since the selection criteria are similar for the two \psg\ samples we use the morphologies of the Y08 (lower redshift sample). We advise the reader to keep in mind the mismatch of the Y08  redshifts and luminosity and we will reconsider the applicability of this comparison when in the discussion.

\subsection{Quasars at $z\sim0.36$}
\label{sec:Samp.QSO}

There exists a well-studied sample of low-luminosity quasars with strong overlap in redshift and luminosity to that of our PSQ sample. This sample consists of 21 broad-lined objects from the SDSS Data Release 7 (DR7) with $r$ model magnitudes $\la19$ and $z\sim0.36$ that all appear in the DR7 Quasar Catalog \citep{schneider10}. Morphologies for the host galaxies of these quasars are available via \emph{HST} ACS imaging \citep[][hereafter T07]{treu07}. Figure \ref{fig:compch4} shows how well matched PSQs and T07 quasars are in $M_r-$redshift space. Given that PSQs harbor quasars somewhat above the quasar/Seyfert dividing line, and the matching in luminosity to the T07 quasars, we consider both PSQs and T07 quasars to be low-luminosity quasars. \citet[][hereafter W06]{woo06} present high-quality Keck-LRIS spectroscopy of the T07 low-luminosity quasars. Despite the existence of the high signal-to-noise ratio Keck spectra, the stellar absorption lines present in many instances are weak, and there was no attempt at explicit stellar population modeling, which would be unlikely to yield robust population properties given the dominant quasar light. We do examine a composite spectrum of this sample constructed from publicly available SDSS data (\S 4), as  well as discuss the results of efforts to determine the stellar populations of more luminous quasars at these redshifts (\S 5).

\section{Analyses}
\label{sec:Anal}

Here we summarize first how the imaging analysis was conducted to provide consistent morphologies of the host  galaxies of our different subclasses.  Second we describe the spectral modeling done to determine the properties of the intermediate-aged stellar population of the PSQs, and explain how we duplicated the fits for the spectra of post-starburst galaxies to make an equivalent set of measurements for comparison. Finally, we describe the measurements and methods made to determine quasar properties, in particular the mass of the central black hole and its Eddington fraction, for the PSQs and the matched quasar sample. 

\subsection{Imaging}
\label{sec:Anal.Img}

Decomposition of galaxy surface brightness profiles enables the discovery of possible mergers and/or interactions. C11 decomposed the HST imaging of the PSQ sample into quasar and host galaxy light. They classified the host galaxy morphologies and characterized signs of merger and/or interaction. Similarly detailed morphological decompositions of the \psg\ and quasar samples are carried out in Y08 and T07. 

Table~\ref{tab:morphch4} gives morphologies and estimates of galaxy disturbance fractions. We obtained the HST imaging for the Y08 \psg\ and T07 low-luminosity quasar samples via the online archive. For consistency we morphologically classify each galaxy in the same manner as our previous morphological study of PSQs (C11). Here we give a brief description of our method of morphological and disturbance classifications. We summarize these in Table~\ref{tab:morphch4}. We refer the reader to C11 for further details regarding these classifications. 

\textit{Spiral} host galaxies contain a disk, arms and/or bars. Some host galaxies have smooth distributions on the large scale but disk-like appearance upon careful inspection thus these galaxies are classified as spirals. \textit{Elliptical} host galaxies have smooth, somewhat featureless light distributions, though they are allowed to show tidal features. For some host galaxies, due to the degree of disturbance, it is difficult to give a classification. Such galaxies are morphologically classified as \textit{Indeterminate}. 
\textit{Disturbed} galaxies show signs of interaction/merger activity such as tidal tails, shells, star-forming knots, and asymmetries. \textit{Undisturbed} galaxies lack asymmetries or signs of interaction/merger activity. 



\subsection{Spectral Modeling}
\label{sec:Anal.Mod}

Spectral modeling enables the characterization of fundamental physical properties of the AGN like black hole mass ($M_{BH}$) and Eddington fraction and host stellar properties such as post-starburst mass and age. C13 fits the PSQ spectra using a $\chi^2$ minimization technique simultaneously modeling the, (i) post-starburst, (ii) narrow-line, and (iii) AGN components. The post-starburst component is described by its age and mass. The AGN component is characterized by a power-law plus broad and narrow emission lines. 


\subsubsection{Post-starburst}


We performed stellar population synthesis modeling of the \psg\ sample with the same post-starburst fitting prescription as used in C13 to fit the PSQ post-starburst population. Specifically, we model the stellar population of the \psgs\ by fitting an instantaneous burst (ISB) which assumes that all stars are coeval with the same chemical composition. The models assume solar metallicity and a \citet{chabrier03} initial mass function \citep[IMF,][]{charlotbruzual07}. The template ages are 56, 75, 100, 133, 177, 237, 316, 422, 562, 750, 1000, 1330, 1770, and 2370 Myr. 

While the spectra are well fit by the models, our model assumptions are simple. Here we consider their effects on our results more closely. In particular, metallicity varies from galaxy to galaxy and within galaxies. For a post-starburst aged $\sim$400 Myr, an uncertainty of $\pm$50 Myr was found when changing the metallicity from solar, $Z\ =$ 0.019, to 0.008 and 0.05. Given the high merger fraction of this sample, dust can be present along the line of sight and affect modeled parameters. For a visual extinction of 0.1 mag, a 422 Myr burst, appears both older and less luminous by $\sim$10\%. Older populations are less affected by dust.

We expect \psgs\ to be composed of young, intermediate, and old stellar populations as corroborated by the Starlight full spectral modeling that include contributions from young, intermediate and old stellar populations, which give flux weighted mean ages of $\sim$ 0.2 dex lower than this study \citep{cidfernandes09}. However, like PSQs, \psgs\ are selected to have dominant intermediate-aged stellar populations. We note that the \psgs, all with EW H$_{\delta}$ $>$ 5 \AA\ (A-star fraction of 40 \%) and many $>$ 7 \AA\ (A-star fraction of 60 \%), represent an extreme population. The spectra are dominated by the post-starburst signatures (i.e., Balmer jump and high order Balmer absorption), therefore we use a single burst to model the intermediate-aged stellar population. We note however that there presumably exists an older underlying population since absorption typical of late-type stars, like G-band and Mg b, are seen in some \psgs. Furthermore, strong [\ion{O}{2}] emission in some of the galaxy spectra suggests the presence of on-going star formation. By making measurements on a suite of mock spectra datasets, we find that the intermediate stellar age and mass can be recovered to better than 10\% as long as the contribution of an older population remains less than 70\% by mass. 

This technique provides good estimates of the age and mass of the intermediate-aged stellar population as well as being consistent method for comparisons to our previous work (C13).

In order to extract host galaxy $u$ and $r$ magnitudes, and thus absolute magnitudes and colors, for both the PSQ host galaxies and \psgs\ we made measurements on the modeled host galaxy contribution to the total spectrum. For each spectral resolution element in the $u$ ($r$) filter, we integrated the flux weighted by the response of the filter. Full details of the processing of the spectra, including correcting for Galactic extinction and shifting to the rest-frame are given in C13. The spectra for both PSQs and G07 \psgs\ are treated identically.

\begin{table*}\scriptsize
\begin{minipage}{140mm}
\caption{Comparisons of Fundamental Physical Properties}
\label{tab:fundch4}
\begin{tabular}{llcccc}
\hline \hline
Sample &  \multicolumn{2}{c}{AGN Properties} & &  \multicolumn{2}{c}{Post-Starburst Properties} \\
\cline{2-3} \cline{5-6}
 & log $M_{BH}$ & $L/L_{Edd}$ & & Age & log Mass \\
 &  ($M_{\odot}$) &   & & (Myr) & ($M_{\odot}$) \\
\hline
PSQs & 8.30$^{+0.05}_{-0.06}$ & 0.047 $\pm$ 0.008 & & intermediate: 1320 $\pm$ 110 & 10.69$^{+0.04}_{-0.05}$ \\
Post-starburst Galaxies & \nodata & \nodata & & intermediate: 960 $\pm$ 100 & 10.97$^{+0.03}_{-0.04}$ \\
$z \sim 0.36$ low-luminosity quasars & 8.36$^{+0.07}_{-0.08}$ & 0.033 $\pm$ 0.004 & &  \nodata & \nodata \\
\hline
\end{tabular}
\end{minipage}
\end{table*}

\subsubsection{AGN}
\label{sec:Anal.AGN}

C13 model the AGN component of the PSQ with a power-law, UV and optical [\ion{Fe}{2}] templates, broad emission lines (\ion{Mg}{2}, H${\beta}$, and H${\alpha}$) and several narrow emission lines. The broad lines are modeled with the sum of two broad Gaussian profiles to deal with asymmetries and broad wings. Prominent narrow lines (i.e., [\ion{Ne}{5}], [\ion{O}{2}], [\ion{Ne}{3}], H${\beta}$, [\ion{O}{3}], and H${\alpha}$) are each fit with a single Gaussian. 

Black hole masses are calculated using virial mass estimates. These estimates employ continuum luminosity and broad line FWHM as proxies for the physical properties of the black hole (i.e., $M_{BH}$ and radius of the broad-line region). C13 calculates the monochromatic luminosity at 5100~\AA\ and $M_{BH}$ for each of the three broad lines, \ion{Mg}{2}, H$_{\beta}$, and H$_{\alpha}$. They give adopted $M_{BH}$ values by taking the median (mean) for three (two) good line measurements. If there is only one reliable value, they adopt this as the $M_{BH}$. 

We note that the relations given for  \ion{Mg}{2} and H$_{\beta}$ are on the same mass scale while the values for H$_{\alpha}$ are higher by a factor of 1.8 \citep{onken04}. For C13, there was no completely self consistent set of scaling relations for \ion{Mg}{2}, H$_{\beta}$, and H$_{\alpha}$. Therefore, the subset of PSQs for which H$_{\alpha}$ was used to calculate the  adopted $M_{BH}$ may have an offset of 0.09 dex, small in comparison to the overall scatter of $M_{BH}$ values \citep[$\sim$ 0.5 dex, see Table 5 of][]{vestergaardpeterson06}.

To estimate a bolometric luminosity, C13 use the monochromatic luminosity at 5100~\AA\ and a bolometric correction, $f$, of 8.1. The Eddington fraction is calculated using:

\begin{equation}
\frac{L_{Bol}}{L_{Edd}} = \frac{f \times L_{5100}}{1.51 \times 10^{38}\ (M_{BH} / M_{\odot})\ ergs\  s^{-1}}.
\end{equation}

The most up-to-date black hole mass estimates for the T07 low-luminosity quasar sample are given in \citet{bennert10}. They use scaling relationships for black hole mass estimates \citep[see Equation 2 of][]{bennert10} with the virial mass scaling of \citet{onken04}. For the line width, $\sigma_{H{\beta}}$, they measure the second moment of the $H{\beta}$ after subtraction of the continuum (windows: 4690$-$4780~\AA\ and 5010$-$5130~\AA) and the [$H{\beta}$] and [\ion{O}{3}] $\lambda$4959~\AA\ narrow lines. Furthermore, they find $L_{5100}$ by extrapolation of the PSF flux of their F110W-band imaging to 5100~\AA\ after subtracting the host stellar continuum. They calculate the Eddington fraction using $L_{Edd}$ $=$ 1.25 $\times$ 10$^{38}$($M_{BH}$/M$_{\odot}$) ergs s$^{-1}$ with bolometric luminosity of $9 \times L_{5100}$, very similar to C13, and resulting in only a very small systematic difference, much smaller than the $M_{BH}$ uncertainty of factors of 3-4. 




\begin{figure}
\includegraphics[width=7.5cm,angle=270]{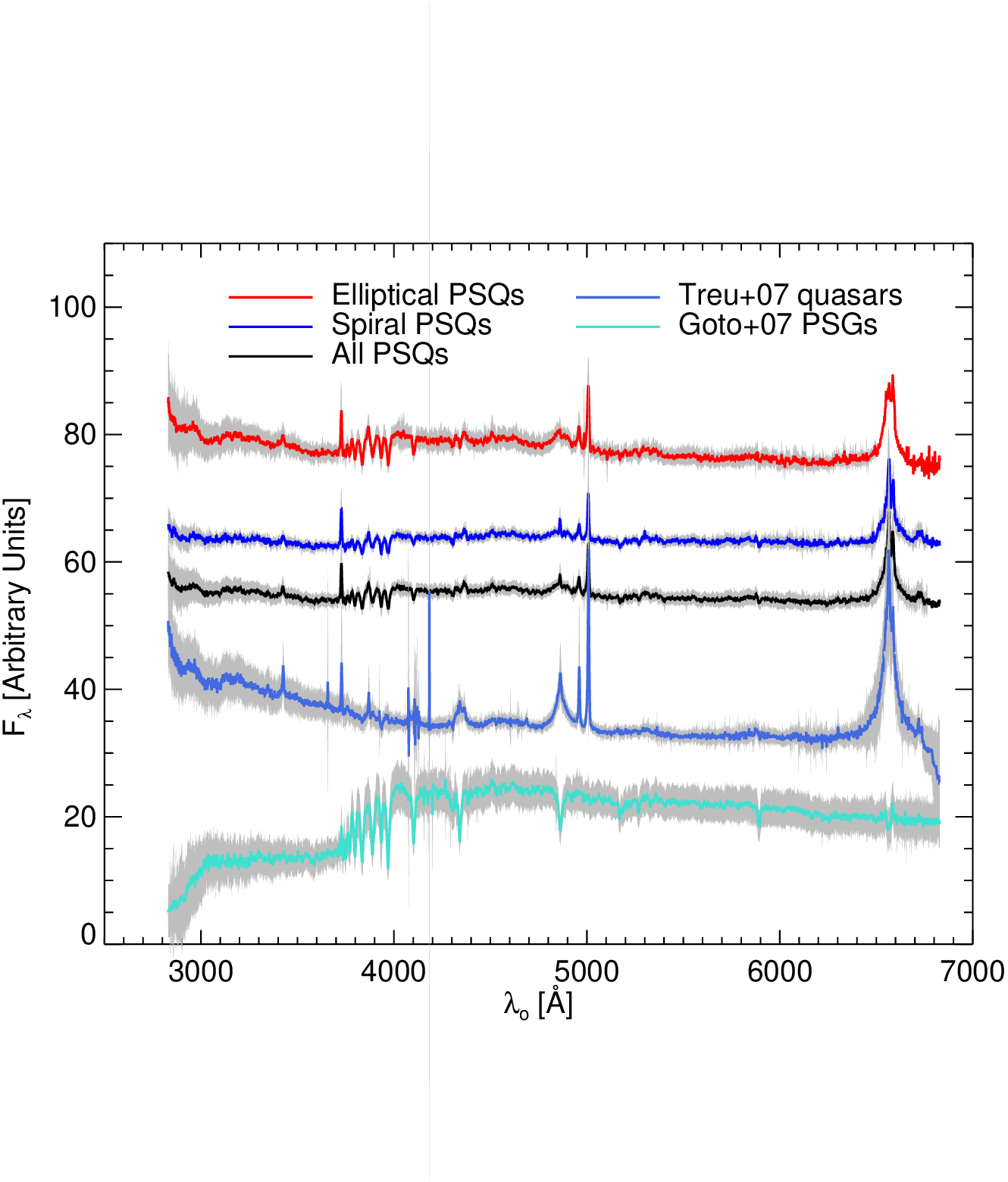}
\caption{Composite spectra and associated RMS spectra of PSQs and comparison samples. From top to bottom the samples are; elliptical PSQs (red), all PSQs (black), spiral PSQs (blue), the T07 low-luminosity quasar sample (light blue) and the G07 \psg\ composite. }
\label{fig:compRMS}
\end{figure}

\begin{figure*}
	\begin{minipage}[!b]{8.cm}
		\centering
		\includegraphics[width=7.8cm,angle=270]{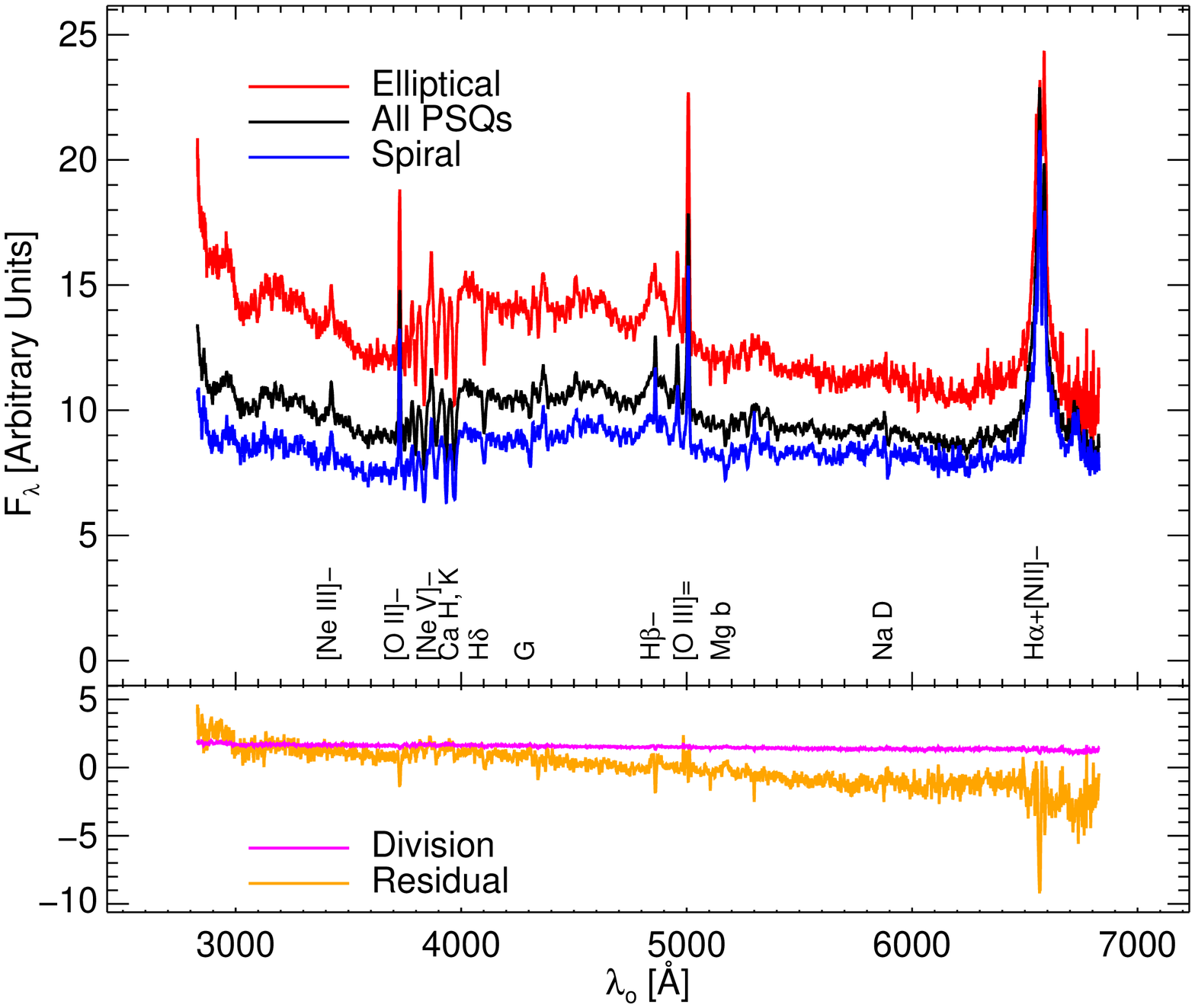}
	\end{minipage}
\hspace{0.4cm}
	\begin{minipage}[!b]{8.cm}
		\centering
		\includegraphics[width=7.8cm,angle=270]{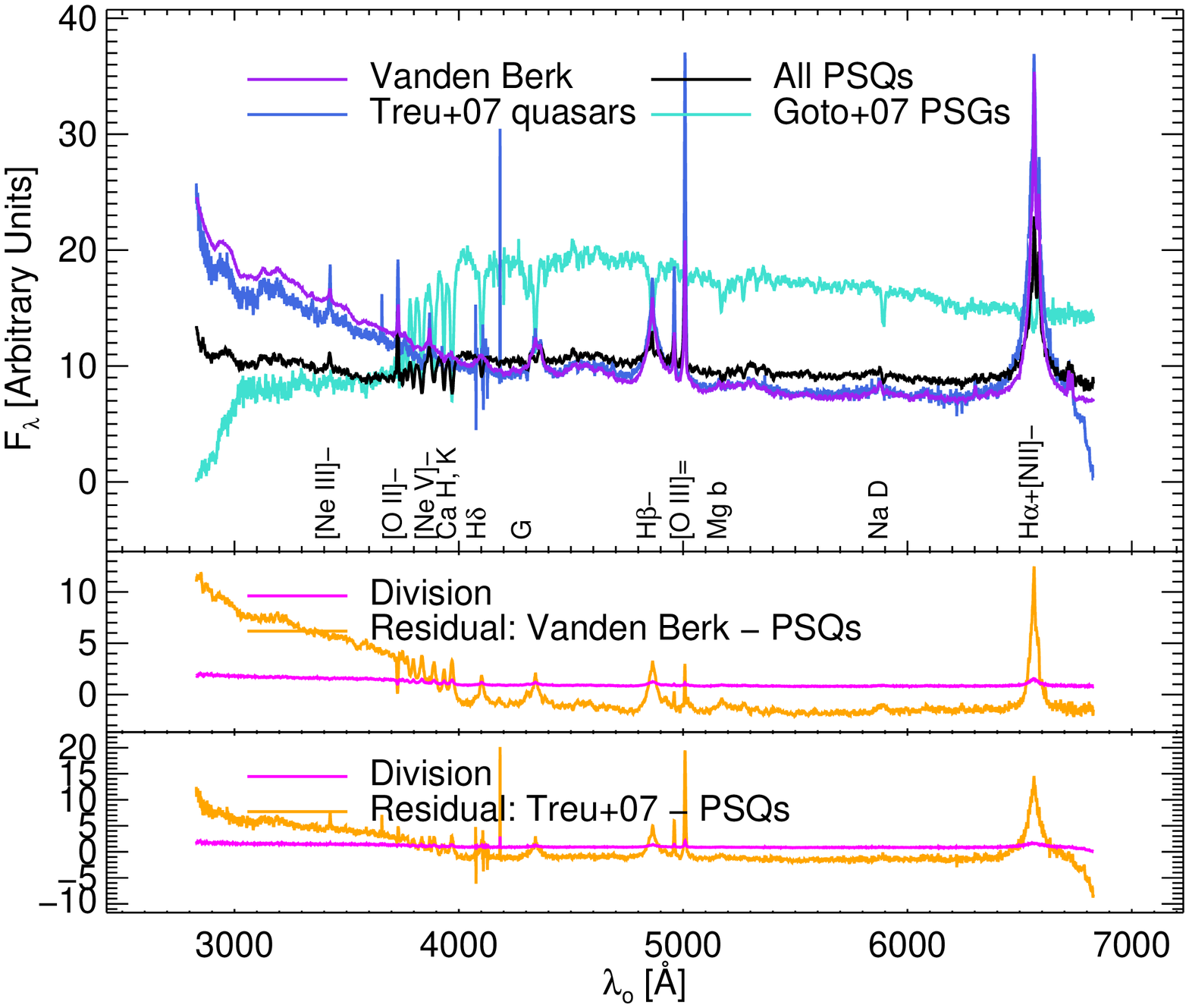}
	\end{minipage}
\caption{{\bf Left:} PSQ composite spectra. The black spectrum is the composite for the total PSQ sample. The red spectrum shows early-type PSQs while the blue spectrum shows the spiral PSQs. Ellipticals are more luminous with a larger starburst.  \label{fig:PSQcomp} {\bf Right:} Composite spectra of the G07 \psgs\ (cyan) and T07 low-luminosity quasars (blue). The V01 composite is plotted in purple. The black spectrum is the composite for the total PSQ sample. Both the \citet{vandenberk01} quasar template and the T07 low-luminosity quasars match well in the red indicating that quasars at this redshift have significant host galaxy contributions. However, the high and low-luminosity quasars are bluer than PSQs. \label{fig:OtherComp}}
\end{figure*}

\begin{figure}
\includegraphics[width=7.5cm,angle=270]{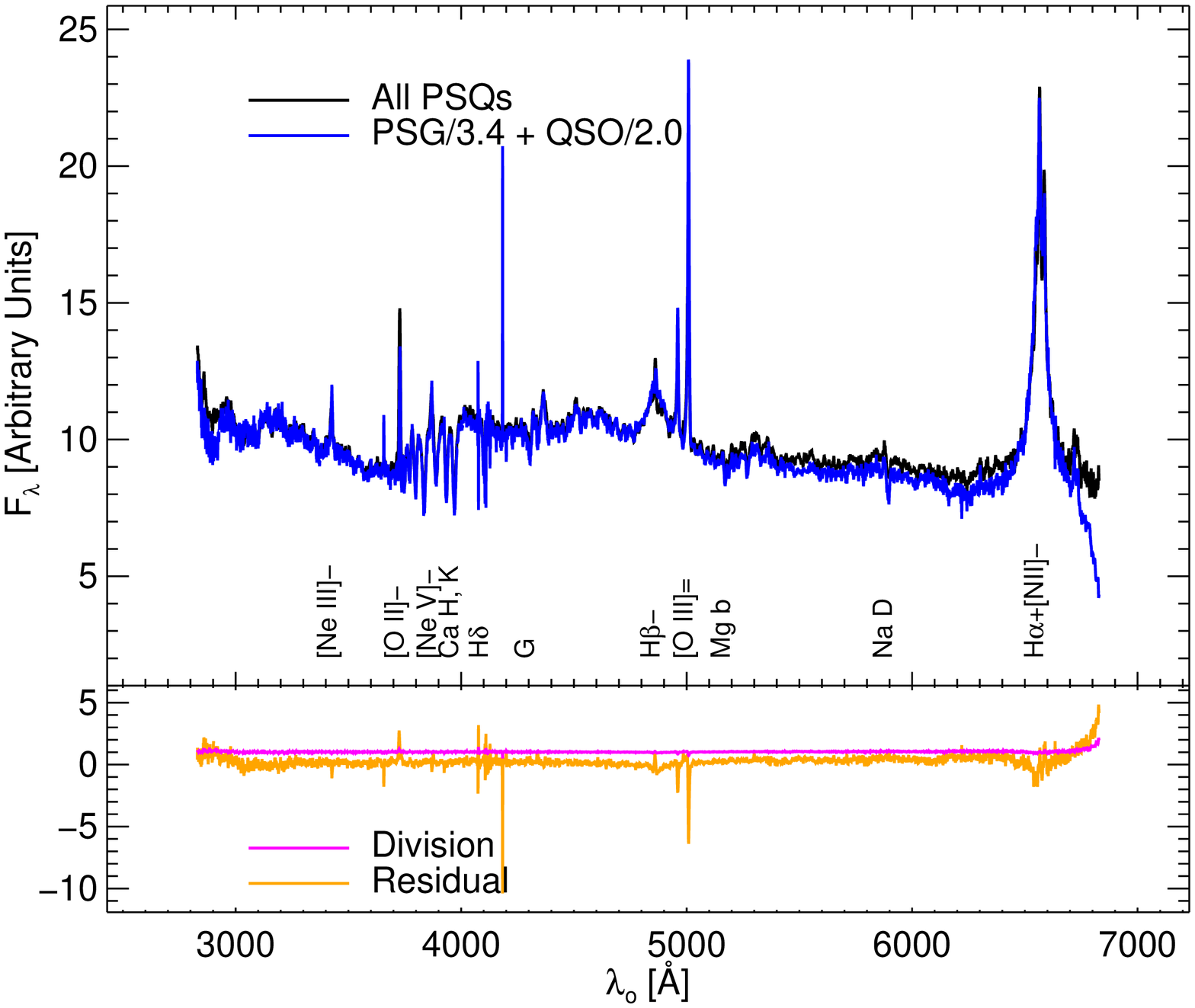}
\caption{Combination of the G07 \psg\ composite (37\% of total integrated light) and the T07 low-luminosity quasar composite (63\% of total integrated light) in blue. The black spectrum is the composite for the total PSQ sample. Residuals between the PSQ composite and the quasar$-$plus$-$\psg\ composite are plotted in both absolute (orange) and relative (magenta) terms. The scaled sum of quasar$-$plus$-$\psg\ is in excellent agreement with the PSQ composite. }
\label{fig:OtherTest}
\end{figure}

\section{Composite Spectra}
\label{sec:Samp.HiLQSO}



We have constructed composite spectra for three samples of PSQs using C13 data:
(i) the full PSQ sample, (ii) the early-type PSQs, (iii) the spiral PSQs.  Using SDSS DR7 spectra,
we also made composite spectra for 
(iv) the G07 \psg\ sample, and (v) the T07 low-luminosity quasar samples. 
For each spectrum in each sample, we corrected for Galactic extinction using the \citet*{schlegel98} maps and the IRAF task \texttt{deredden} which uses the \citet*{cardelli89} extinction curves. We converted the spectra to rest frame using the IRAF \texttt{dopcor} task and the SDSS pipeline redshifts. We made the combinations in the same manner for each sample by scaling each individual spectrum by its mean and then average combining the spectra for the common wavelength range of 2830 to 6829~\AA\ with 1~\AA\ sampling. To characterize the spectrum-to-spectrum difference for each sample we first scale each contributing spectra by the sample mean and then calculate the root-mean-squared (RMS) spectra. Figure~\ref{fig:compRMS} shows the composite spectra for each sample along with the sample's associated RMS spectra.

The left panel of Figure~\ref{fig:PSQcomp} shows the spectral composites for the full PSQ sample as well as the early-type and spiral subclasses. The figure demonstrates that the early-type PSQs are more luminous. C11 find that when compared to spiral hosts the early-type PSQs have greater PSF (AGN) luminosities (p-value, the probability of observing such a difference of 0.77\%). C13 find that the total integrated light in the spectra from 3000 to 6000 \AA\ is higher for early-type hosts than spirals (p-value of 0.039\%). Furthermore, when ranked by luminosity, 9 of the 11 most luminous spectra are early-type hosts.

The right panel of Figure~\ref{fig:OtherComp} shows the spectral composites for the full PSQ sample as well as the G07 \psgs\ and T07 low-luminosity quasars. Additionally, we plot the composite SDSS quasar spectrum from \citet{vandenberk01}. Note that host galaxy contributions are included in all spectra. With no scaling between the composites, it is apparent that the \psgs\ are more luminous in the red. When compared to PSQs, both the \citet{vandenberk01} quasar template and the T07 low-luminosity quasars are only slightly less luminous in the red indicating that most quasars at these luminosities likely have similar host galaxy contributions from an older stellar population. However both high and low-luminosity quasars are bluer and as a result slightly more luminous, consistent with the PSQs having a significant contribution from a post-starburst stellar population that is overall red compared to typical quasar colors.

We want to comment more on the T07 composite spectrum.  The \ion{Ca}{2} K $\lambda$3933 absorption line is significantly detected (EW = $1.7 \pm 0.04$ \AA). \ion{Ca}{2} H $\lambda$3968, blended with H$\epsilon$ $\lambda$3970, and other high-order Balmer lines are undetected (EW $\leq$ 0.04 \AA). This is suggestive that the Balmer absorption lines are intrinsically weak and/or filled in with emission, indicating that these normal low-luminosity quasars are hosted in galaxies with a significant contribution of an old stellar population ($\ga5$ Gyr) and not dominated by a post-starburst. 

We experimented to see if we could reproduce the PSQ composite spectrum with a combination of the composite spectra from the other classes. Figure~\ref{fig:OtherTest} shows that the scaled superposition of G07 \psg\ and a T07 low-luminosity quasar composites are remarkably similar to the average PSQ spectrum. The \psg\ composite contributes 63\% of the total flux while the low-luminosity quasar composite contributes 37\%. We show residuals between the PSQ composite and the quasar$-$plus$-$\psg\ composite in both absolute and relative terms. The reduced-$\chi^2$ is 0.4, for $>4000$ degrees of freedom indicating a very good match. We conclude that the PSQ average spectrum is largely consistent with a scaled sum of \psg\ and quasar composite spectra at similar z.

It is important to note that since the PSQs were selected using an apparent magnitude cut, the total luminosity reflects contributions from both stellar and AGN components. In order to detect an object as a PSQ, the post-starburst population and AGN luminosity must be of comparable strength. C11 find for the PSQ sample that the average contribution from the AGN to the total light in the HST/ACS F606W filter is $\sim30$\%. Furthermore, the integrated light in the spectra from 3000 to 6000 \AA \ of the AGN and post-starburst components are very nearly the same. These scaling coefficients give us important clues about our selection biases. In order to completely understand these biases and to improve future sample selection, it is necessary to run our selection on mock datasets that represent varying contributions of AGN and post-starburst \citep[][ \textit{in preparation} ]{cales15}.

For the PSQ composite and quasar$-$plus$-$\psg\ composite, there is little to no difference between AGN power-law slopes or broad-line features. Additionally, the post-starburst shapes are comparable and there are no residuals between the Balmer absorption line series. The most notable differences appear to be between the narrow lines where the PSQ composite has more flux in [\ion{O}{2}], H$_{\beta}$ and [\ion{S}{2}] lines and less flux in the [\ion{Ne}{3}] and [\ion{O}{3}] lines. As found in C13, there is a large range in contributions of narrow lines so the difference could be due to varying contributions of ongoing star formation.


\section{Results and Discussion}
\label{sec:Res}


We characterized the ages and masses of the host stellar populations, and the black hole masses and Eddington fractions of the AGN components in our samples. Table~\ref{tab:fundch4} summarizes these properties.  Additional results are integrated into the discussion below.

There is mounting evidence that galaxy evolution is at least bimodal (if not multi-modal). Luminous early-type galaxies are thought to be the product of major-mergers. Through simulations, major galaxy mergers have been found to be efficient at driving gas to the galaxy center \citep{barneshernquist91}, where it can be used as fuel for both intense circumnuclear star formation and black hole growth. In turn radiative and/or mechanical feedback from the accreting BH can inhibit both star formation and its own fueling \citep{dimatteo05, springel05, hopkins06}. These galaxies went through a brief but catastrophic quenching phase and are observed to have little to no ongoing star-formation along with intermediate to old-aged stellar populations. Numerical simulations have successfully reproduced the physical properties of elliptical galaxies and bulges through major mergers of gas-rich disk galaxies \citep{granato04, dimatteo05, hopkins06}. Observational evidence of mergers are seen in the form of shells and tidal tails in elliptical galaxies \citep{malincarter80, schweizer80, canalizo07}. Most spiral galaxies have not undergone such a violent event and with their plentiful gas reservoirs can maintain low-to-intermediate levels of star formation giving rise to more complex star formation histories \citep{hopkinshernquist09}.

To explore the role PSQs play in galaxy evolution we compare and contrast the morphologies and fundamental physical properties of galaxy types which appear to be most closely related to PSQs (i.e., \psgs\ and quasars). For example, comparing the stellar population ages of PSQs to \psgs\ may lead to a clearer picture of where PSQs fall in an evolutionary scenario of galaxies. Furthermore, a comparison of PSQs to quasars of similar redshift and luminosity may lead to gained insight into the black hole demographics of PSQs.

\subsection{Morphology}
\label{sec:Res.Morph}

The host morphologies of the PSQs are heterogeneous and split between early-type and spiral. Y08 \psg\ morphologies are varied although their bulge-to-total light decompositions are consistently early-type. Additionally, many of the \psgs\ are markedly dusty with a high fraction of disturbances (see Table~\ref{tab:morphch4}). However, the morphologies of the low-luminosity quasar sample and light profile decompositions are skewed towards spiral type with a lower fraction of disturbances. These results are consistent with the idea of bimodal galaxy evolution. At this redshift \psgs\ and low-luminosity quasars are at either end of the bimodal distribution of galaxy evolution with PSQs nestled in the middle bridging the gap. Above a certain luminosity ($L_{Bol} > 10^{46}$ ergs s$^{-1}$), disturbed and post-merger ellipticals dominate, below largely undisturbed spirals dominate.

There may be a number of selection or observational effects that may contribute to the observed morphologies and disturbance fractions. With longer exposure times, spiral structure and signposts of interaction become more apparent. The luminosity cut of the Y08 \psgs\ ($M_r \sim -21.8$) leaves the sample incomplete and likely skewed towards more massive systems, making them more similar to G07 \psgs. To ensure little ongoing star formation, post-starburst selection algorithms generally place a limit on nebular emission ([\ion{O}{3}] and H$_{\alpha}$). This has the effect of biasing post-starburst galaxies to have early-type morphology. Given that disturbed elliptical galaxies are more luminous than their non-disturbed counterparts \citep{darg10a, darg10b}, it is hard to imagine the even more luminous G07 systems to not be disturbed merger products.  Furthermore, the PSQs necessarily have both high-luminosity AGN and a significant contribution of continuum from the host galaxy (integrated AGN and starburst luminosities from 3000 to 6000 \AA\ of $L_{AGN}> 10^{43.15}$ ergs s$^{-1}$ and $L_{SB} > 10^{43.15}$ ergs s$^{-1}$). Increasing the AGN luminosity would thus swamp out contributions from host stellar populations. Consequently, while the host galaxy may be quite luminous, PSQs contain Seyfert to low-luminosity quasars and miss the highest luminosity quasars. 

The most careful imaging studies of higher luminosity quasars \citep{canalizo07, bennert08} do in fact show that they are hosted in disturbed ellipticals, contrary to earlier, shallower investigations that missed the disturbances. 
\citet{canalizo07} and \citet{bennert08} estimate that the contribution of light in fine structure (disturbance features) for luminous quasars is between 6 and 13\%. Similarly, we find that the PSQ disturbance features account for 3-28\% of the total light with a mean of $\sim$10\%. 


\begin{figure}
\includegraphics[width=7.5cm,angle=270]{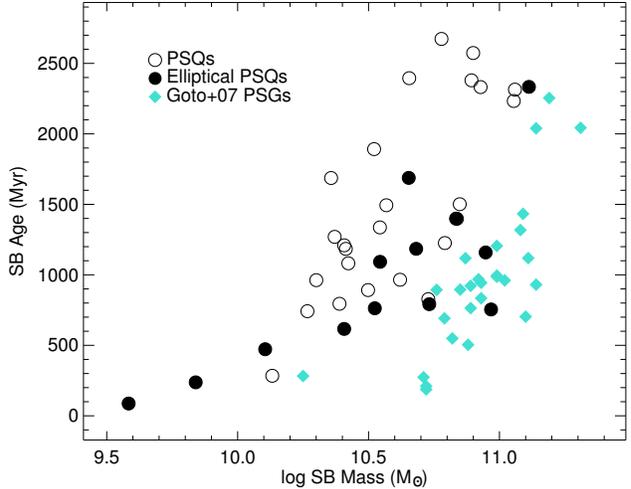}
\caption{Post-starburst age versus mass for the stellar population models fit to the spectra of PSQs (black circles) and G07 \psgs\ (cyan diamonds). For a given overlapping mass there exist \psgs\ that could evolve to become PSQs. Average errors for post-starburst age (41 Myr) and mass (4 $\times$ 10$^8$ M$_{\odot}$) given by the spectral modeling are on par with the size of the points.}
\label{fig:SBmassageGoto}
\end{figure}

\begin{figure}
\includegraphics[width=7.5cm,angle=270]{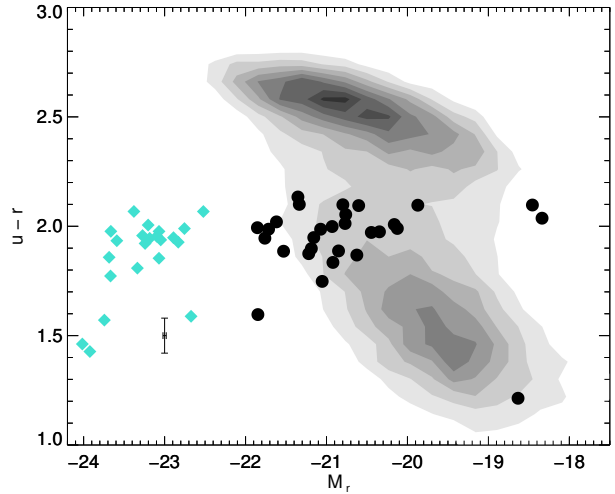}
\caption{Color-magnitude diagram of PSQ hosts (black circles) and G07 \psgs\ (cyan diamonds) over the local SDSS galaxy population (contours). $u - r$ colors are measured for PSQs and \psgs\ from the starburst component of the spectra after correcting for Galactic extinction and shifting to the rest frame. The local SDSS galaxy $u - r$ colors are corrected for Galactic extinction. PSQs and G07 \psgs\ have green colors indicative of an intermediate-aged stellar population, but more luminous than typical galaxies of similar color. The error bars shown in the bottom left represent maximum errors for the $u-r$ color term and $M_r$ (0.08 and 0.03, respectively).}
\label{fig:colormagGoto}
\end{figure}

\subsection{Post-Starburst Stellar Properties}
\label{sec:Res.Der.SB}

Given the typical duty cycles of AGN ($\sim10^8$ Myr or less) in addition to the lifetime of post-starburst spectral signatures ($\sim$1-2 Gyr), PSQs may evolve from \psgs. If so, then \psgs\ should include objects with post-starburst populations of similar masses yet younger ages. We test whether PSQs are evolved from \psgs\ by comparing the post-starburst ages determined from the spectral modeling. 

Figure~\ref{fig:SBmassageGoto} plots the starburst ages of PSQs and \psgs\ from our spectral fitting.  The cut offs for both PSQs and \psgs\ at older ages and lower masses are primarily a result of selection effects, which are different for the two types.  Therefore not too much should be inferred from the relative lack of overlap -- a lower S/N cut by G07 would have resulted in more \psgs\ overlapping the PSQs at older ages and lower masses.  The cut off at the youngest ages and largest masses is real, however, and represents the population demographics as well as the likelihood that the youngest systems are likely dust-enshrouded and classified as ULIRGs.  It is of interest that the cut offs for the PSQs and \psgs\ differ in the lower right corner of the figure, which may indicate that the youngest starbursts of a given mass are seen as \psgs\ rather than PSQs, which would make the AGN activity something that lags the youngest \psg\ phases \citep[see e.g.,][]{wild07}.  Between the selection effects and the apparently real preference for \psgs\ among the youngest post-starbursts for a given mass, the \psgs\ are systematically more luminous than the PSQs in our comparison, as seen in the color-magnitude diagram of Figure~\ref{fig:colormagGoto}.  PSQs and G07 \psgs\ are high luminosity objects with green colors, indicative of intermediate-aged stellar population, yet the post-starburst galaxies are more luminous as a results of having younger ages and a shallower selection. We also note that there may be an over-density in massive post-starburst galaxies in comparison to PSQs. Thus, we conclude that some but not all post-starburst galaxies may evolve to a PSQ phase. Indeed, the most massive post-starburst galaxies may be so efficient at forming stars that the phase can run away, starving the AGN of what otherwise might have been $M_{BH}$ growing fuel \citep{mitton92}. Furthermore, the AGN of the PSQs, once turned on, may truncate further star formation resulting in the observed smaller post-starburst stellar mass (and luminosity).

The results are clear: there exists a population of \psgs\ with similar mass, yet younger stellar populations than PSQs, suggestive of a delay in AGN-triggering following a merger or major interaction that leads to a starburst. It seems that observationally and theoretically there is growing consensus regarding the AGN fueling in the context of merger driven evolutionary scenarios in which bursts of star formation happen early and coincide with the close approach passes of the two nuclei, while the AGN is triggered later on in the evolution of the merger around coalescence \citep{wild07, schawinski10_2mech, vanwassenhove12, alexanderhickox12, canalizo13}. At the sub-kpc scales, competing effects (e.g., inflow, outflow, star-formation) cause the gas to pile up \citep{alexanderhickox12, fathi13}. At any rate, this suggests that getting fuel all the way down to the BH is not straight forward and takes time. Our current view of merger triggered AGN fueling is one in which torques from the merger are efficient drivers for bringing gas down to a critical distance from the BH ($\sim$100pc). Then a series of small-scale internal instabilities (stochastic fueling) finally brings the gas to the BH, where the time it takes the gas to reach the BH may be comparable to or longer than the AGN duty cycle.



\subsection{Stellar Populations in Hosts of Luminous Quasars}
\label{sec:Res.Der.luminous}

The composite quasar spectrum of T07 did not show evidence of post-starburst stellar populations, and the HST images showed relatively undisturbed spiral hosts, in stark contrast to many of our PSQs.  In recent years, significant work has been done to determine the stellar populations of luminous quasars -- more luminous than our PSQs and the T07 sample -- at these redshifts.

Recently \citet{canalizo13} presented an analysis of off-nuclear host galaxy spectra of a sample of luminous quasars at redshifts of about 0.2.  In that investigation, 14/15 objects had spectra inconsistent with simple, older stellar populations, and required significant contributions from post-starburst populations.  These objects also have elliptical host galaxies showing various morphological signatures of mergers that likely triggered activity.  In fact, \citet{canalizo13} claim that that their results are the same as the results of C13.  To verify whether the intermediate-age stellar populations in PSQs and the classical QSO host galaxies of \citet{canalizo13} are consistent with being drawn from the same parent population we employ the nonparametric Kolmogorov-Smirnov test. The PSQs and the classical QSO host galaxies with average starburst ages of 1320$\pm$110 Myr and 1430$\pm$180 Myr, respectively, are consistent with having the same starburst ages (D$=$0.300, P-value$=$0.261).  The host galaxies of PSQs and luminous quasars do indeed possess essentially the {\em same} post-starburst stellar populations.  This is in stark contrast to the comparison with the T07 low luminosity quasar composite, for which significant intermediate-aged stellar populations appear to be the exception, not the norm.

The \citet{canalizo13} results are consistent with a number of other recent investigations \citep{letawe07, jahnke07, wold10}. Deeper imaging and careful analysis seem to be converging on a picture in which luminous quasars have elliptical galaxies showing evidence of mergers and significant younger stellar populations that are inconsistent with passively evolving red elliptical galaxies.  Our sample of PSQs may be more closely related to more luminous quasars than those of similar luminosity (T07).  A viable hypothesis is that the PSQs may in fact be transitioning into, or out of, a higher luminosity phase, in which the post-starburst is easily hidden in the glare of a more extreme quasar, but easily selected in the transition.

\begin{figure}
\includegraphics[width=7.5cm,angle=270]{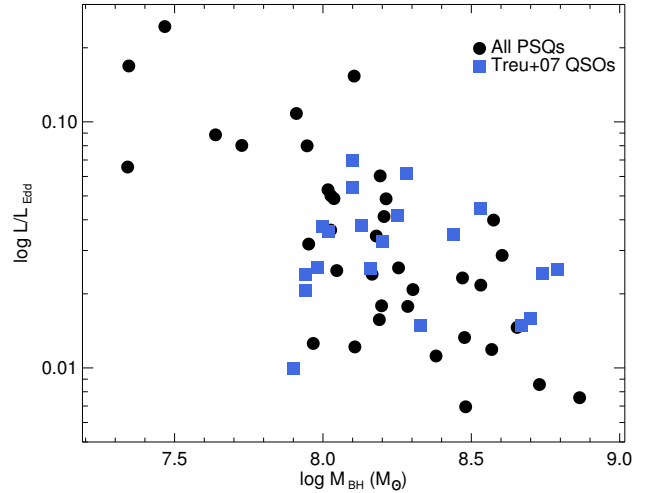}
\caption{$M_{BH}$ versus $L/L{Edd}$ for the PSQ sample (black circles) and the T07 low-luminosity  quasar sample (blue squares). }
\label{fig:AGNcomp}
\end{figure}

\begin{figure*}
	\begin{minipage}[!b]{8.cm}
		\centering
		\includegraphics[width=7.9cm,angle=270]{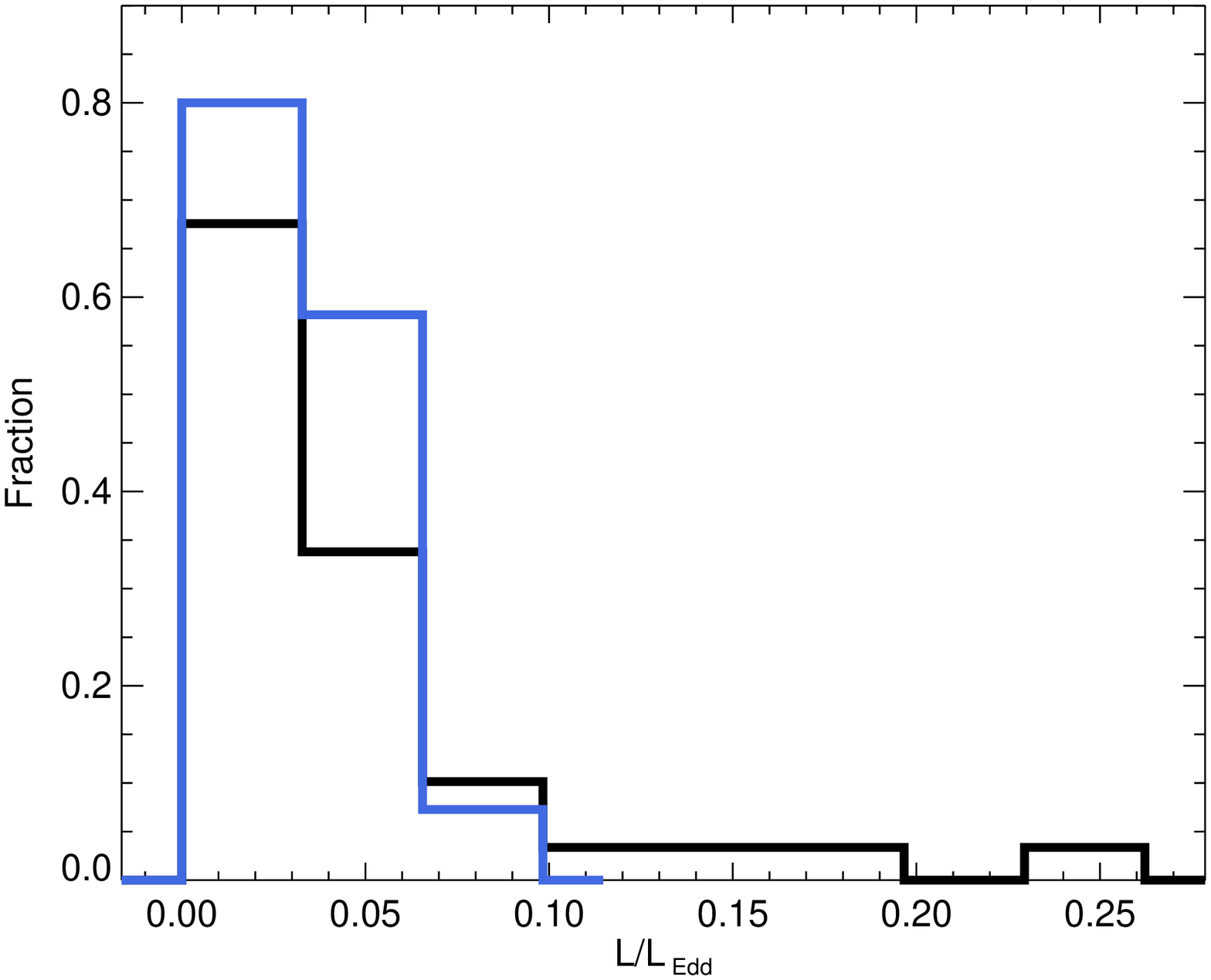}
	\end{minipage}
\hspace{0.4cm}
	\begin{minipage}[!b]{8.cm}
		\centering
		\includegraphics[width=7.9cm,angle=270]{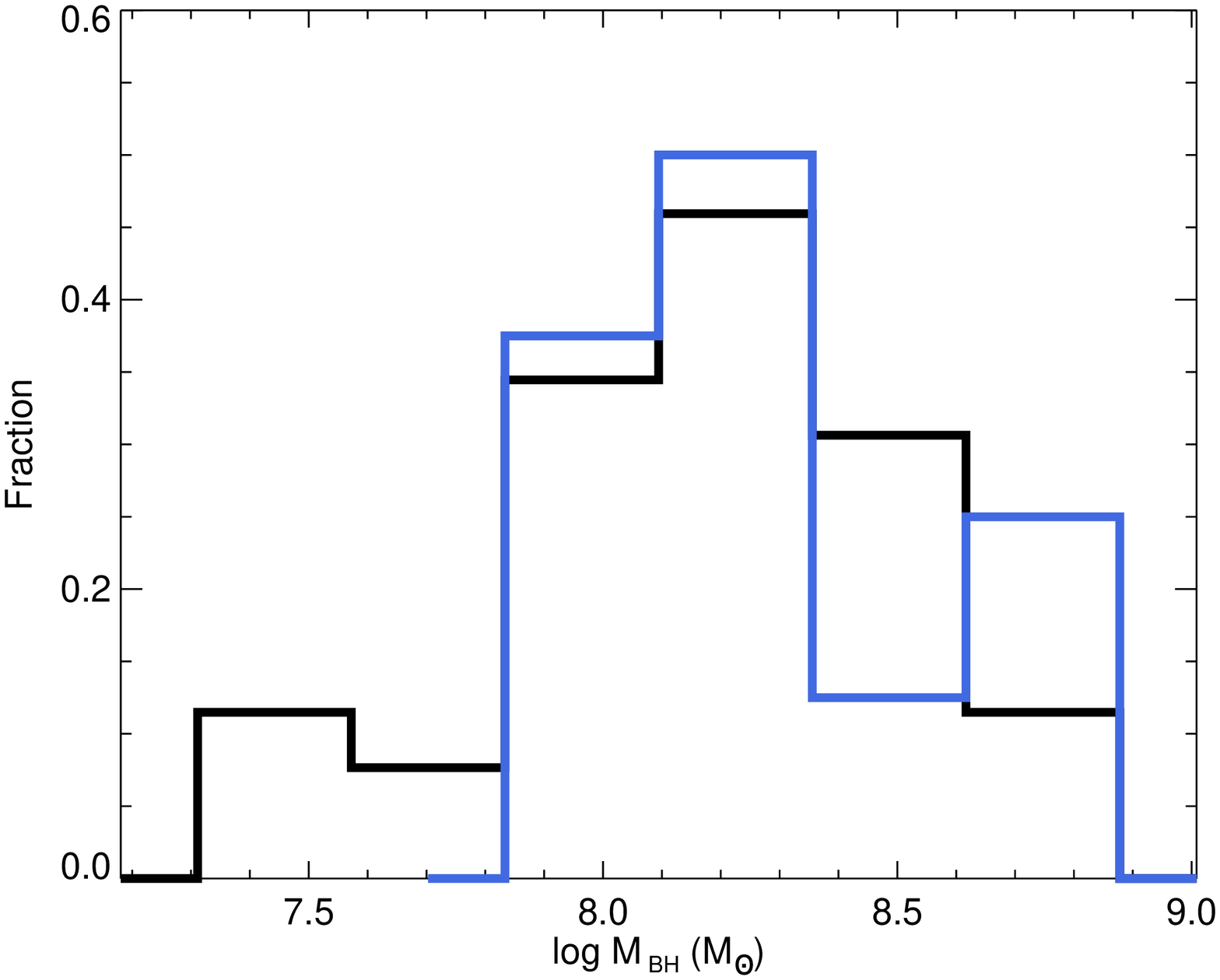}
	\end{minipage}
\caption{ {\bf Left:} $M_{BH}$ histogram for the PSQ (black) and T07 low-luminosity quasar (blue) samples. The $L/L{Edd}$ distributions for the two samples are not statistically different. \label{fig:histLfrac} {\bf Right:} $M_{BH}$ histogram for the PSQ (black) and T07 low-luminosity quasar (blue) samples. The $M_{BH}$ distributions for the two samples are not statistically different. \label{fig:histMbh}}
\end{figure*}


\subsection{AGN Properties}
\label{sec:Res.Der.AGN}

Figures~\ref{fig:AGNcomp} and \ref{fig:histMbh} show the distribution and histograms of $M_{BH}$ and $L/L_{Edd}$. Based on results from the nonparametric Kolmogorov-Smirnov tests we cannot reject the null hypothesis that PSQs and the T07 low-luminosity quasars are drawn from the same population for both $M_{BH}$ (D$=$0.132, P-value$=$0.966) and $L/L_{Edd}$ (D$=$0.213, P-value$=$0.538). The PSQs and the T07 low-luminosity quasars are consistent with having the same $M_{BH}$ and Eddington fractions (few $\times 10^8\ M_{\odot}$ and a few percent, respectively). This provides further evidence that the AGN contribution to PSQs are typical of `normal' quasars at this redshift. However, PSQs occupy the low Eddington fraction high-mass end of the \citet{borosongreen92} radio-quiet low-redshift quasars which may indicate that the AGN of the PSQs may not be at their peak activity.


\subsection{Merger Driven Nuclear Activity}
\label{sec:Res.Der.Merger}

\begin{figure*}
\includegraphics[width=15cm,angle=0]{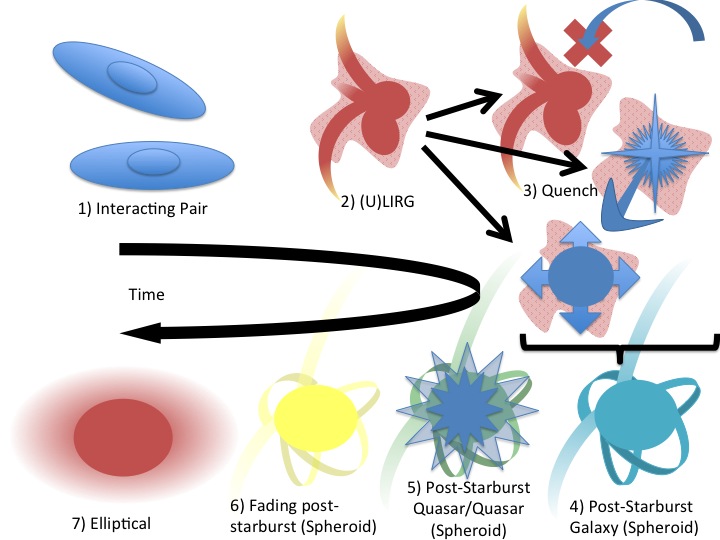}
\caption{Diagram outlining our scenario for merger triggered PSQs. Close approach galaxy-galaxy passages induce massive bursts in star formation. ULIRGs are ubiquitously merging systems harboring heavily dust enshrouded and intense starbursts. A quenching phase (halo, SN or AGN triggered) halts star formation. The galaxy emerges as a post-starburst. Once gas has made it all the way to the BH an episode of quasar activity ensues. Finally after $\sim$1-2 Gyr a `red and dead' elliptical remains.}
\label{fig:merger}
\end{figure*}

Major-merger evolutionary scenarios predict a sequence of transitioning galaxies, with particular observable properties. Figure~\ref{fig:merger} illustrates the galaxy transformations in the context of merger driven evolution and making a few improvements on the current view of merger triggered galaxy evolution in order to reflect new findings regarding the how star formation is quenched, the most up-to-date merger simulations and the timing/triggering of PSQs the following describes the proposed sequence. (1) At the largest scales as two galaxies are coming together, they interact. Several thousand galaxy pairs and mergers have been discovered in the SDSS \citep{ellison08, darg10a}. (2) As the merger continues, galaxy-galaxy dynamical interactions can funnel fuel into the center of the galaxy triggering star formation. Merger remnants are typically ultra-luminous infrared galaxies (ULIRGs) enshrouded in dust, many of which have buried AGN \citep[e.g.,][ and references therein]{sandersmirabel96}. (3) Then, shutting off the gas supply quenches the star formation. There are several mechanisms that could be responsible for quenching star formation. The system could simply have converted all (most) of the gas made available through the merger to stars in an intense starburst episode. Galactic-scale winds could also play a part. With an intense bout of star formation come intense supernova winds \citep{heckman87, tremonti07, diamondstanic12}. Furthermore, feedback from a quasar episode could also remove fuel for star formation from the galaxy \citep[see][ for review]{alexanderhickox12}. (4,6) The remnant evolves to a post-starburst or E$+$A galaxy, a couple hundred million years after a major burst in star formation. \Psgs\ tend to have elliptical morphologies with signs of disturbance \citep{yang08}. (5) Recent simulations by \citet{vanwassenhove12} show that massive  bursts of star formation occur during the closest approaches, while the AGN is activated further on in the merger process at coalescence.  At some time during the post-starburst phase and through the combined efforts of (i) large scale gravitational torques and (ii) a series of small scale internal instabilities, the cold gas can loose 
$\sim$99.99\% of its angular momentum and finally fuel a quasar for a time even shorter than its journey to the supermassive black hole \citep[for observational studies of PSQs see][]{cales11,cales13}. The quasar phase ($\sim$10$^{7-8}$ yr) is short in comparison to the lifetime of post-starburst stellar populations \citep[0.3-3 Gyr,][]{falkenberg09,wild09}. (7) Eventually after a couple of billion years (Gyr), a spheroidal galaxy emerges with properties that match those of local ellipticals. 




\subsection{Secular Black Hole Growth and Host Galaxy Transitions}
\label{sec:Res.Der.Merger}

In the merger picture the nuclear activity has a common trigger. However, triggering stochastic accretion of gas onto the black hole can be unrelated to properties of the host in which it lives. Especially for low to moderate luminosity systems, AGN activity might be an episodic phenomenon, so the low luminosity spiral post-starburst quasars might represent a chance snapshot of dual AGN-starburst activity. However, bars are efficient at driving gas inwards, which could fuel both star formation and a BH \citep[see ][ for a review]{jogee06}. Observationally, the bar fraction is high for narrow-line Seyfert 1s \citep{crenshaw03}. High speed fly-by encounters (harassment) could also cause gas to funnel to galactic nuclei, however, harassment is generally a cluster phenomena. Furthermore, to explain disky PSQs, this requires some fine tuning of the approach conditions such that the disk remains intact. 

Still one needs to explain the (slow) quenching of star formation in these systems. The system could be unplugged from the cosmic supply of gas such that gas can no longer accrete from the surrounding IGM \citep{lilly13}, for example, the halo (bulge) could have grown large enough to support against further star formation \citep{kauffmann03, cattaneo06, martig09}, or the AGN could be injecting energy into the halo, keeping the gas warm so that it cannot collapse and form stars \citep[i.e., `maintenance mode AGN', e.g.,][]{best05}. 


\section{Summary}
\label{sec:Summ}

High-spatial resolution HST imaging and high-quality spectroscopy of samples of PSQs, \psgs\ and quasars detail the differences and similarities of their morphologies and optical properties. We found the following:


\begin{itemize}
\item \textbf{Modes of evolution:} \Psgs\ and high-luminosity quasars ($L_{Tot} >$ 10$^{43.85}$ erg s$^{-1}$) are disturbed early-type galaxies consistent with a major-merger mode (elliptical mode in C11) of evolution. Lower luminosity quasars ($L_{Tot}\ <$ 10$^{43.95}$ erg s$^{-1}$) with spiral hosts, show less signs of disturbance consistent with a Seyfert/secular/downsized mode (spiral mode in C11) of evolution. PSQs have a variety of morphologies, including elliptical, spiral, disturbed, undisturbed, interacting, and singular. This indicates that both dynamical interactions and secular evolution appear to play important roles in the evolution of luminous active galaxies while $z < 1$.
\item \textbf{PSQ Composites:} A morphological comparison between PSQ spectral composites confirms findings from C11 and C13 that early-type PSQs have younger, burstier stellar populations ($\langle$SB Age$\rangle \sim$960 Myr) and more luminous AGN ($\langle L_{Tot}\rangle \sim$ 10$^{44.08}$ erg s$^{-1}$) providing strong evidence for a merger-driven evolutionary connection between the starburst and AGN for the early-type PSQs. 
\item \textbf{Composite comparisons:} A quasar$-$plus$-$\psg\ composite, when scaled appropriately, compared to a PSQ composite is in excellent agreement, recovering the main features of PSQs, and is highly suggestive that there is indeed a link between PSQs, \psgs, and quasars at this redshift. 
\item \textbf{Post-starburst galaxy comparisons:} Like PSQs, \psgs\ show a high disturbance fraction and varied morphologies though they are skewed toward early-type morphology. Post-starburst galaxies have a similar colors and masses (log SB Mass: 10.7$-$11.2 $M_{\odot}$) yet younger post-starburst populations than PSQs as a whole. When only compared to the early-type PSQs, \psgs\ have roughly the same age populations ($\langle$SB Age$\rangle \sim$960 Myr). We show that PSQs are consistent with being evolved \psgs.
\item \textbf{Quasar comparisons:} The quasars matched in redshift and luminosity to the PSQs tend toward spiral morphology with lower disturbance fractions. The PSQs and the luminosity and redshift-matched T07 quasars are consistent with having the same $M_{BH}$ and Eddington fractions ($\sim$ 10$^{7.5} - $10$^{9.0}$ $M_{\odot}$ and $\sim$1\%$-$10\% Eddington). The stellar continuum contribution seen in the low-luminosity quasars is consistent with old stellar populations ($\ga5$ Gyr). However, comparisons of PSQ stellar populations and those determined for more luminous quasars indicate greater similarities, at least with the PSQs that appear to be merger products.
\end{itemize}

Galaxy mergers appear capable of triggering AGN activity via dynamical interactions that also fuel intense bursts of star formation. Circumnuclear star formation occurs early in the merger corresponding to close passes of the merging nuclei. From this extended circumnuclear region,  a series of small-scale internal instabilities eventually bring gas all the way down to the BH. The AGN activity occurs later during the evolution of the merger around coalescence. Observationally, \psgs\ emerge some several 100 Myr after the first close approach coincident with the observed/predicted delays in AGN/quasar activity. In this context, the highest-luminosity AGN (quasars) are hosted by \psgs\ and PSQs are observable when the AGN is not swamping the light of the host (i.e., transitioning into or out of their most luminous phase).

Secular evolution plays a significant role in the local universe ($z \la 1$). Spiral galaxies, with their plentiful gas reservoirs, can maintain low-to-intermediate levels of star formation and complex star formation histories. Stochastic accretion of gas can cause episodic AGN activity at low levels, thus in a sample of Seyferts, one might expect to find a range in star formation histories and the low-luminosity spiral PSQs may represent a chance snapshot of simultaneous AGN-post-starburst activity.

Acknowledgements: We thank the anonymous referee for their helpful comments. SLC was supported by ALMA-CONICYT program 31110020.


\bibliographystyle{./mn2e}
\bibliography{./AllCat}


\label{lastpage}
\end{document}